\documentclass[aps,prl,reprint,floatfix,footinbib,notitlepage,grouppedaddress,showpacs]{revtex4-2}
\usepackage{graphicx,graphics,times,bm,bbm,bbold,amssymb,amsmath,amsfonts,mathtools,dsfont,xcolor,color,cancel}
\usepackage[colorlinks=true,citecolor=purple,linkcolor=purple,urlcolor=purple]{hyperref}
\usepackage{soul} 
\setstcolor{red} 

\usepackage{braket}

\newcommand*\diff{\mathop{}\!\mathrm{d}\hspace{0.3pt}}
\newcommand{\dt}{\diff t}

\makeatletter
\def\@bibinfo#1#2{%
  \ifx#1volume\textnormal{#2}%
  \else\ifx#1issue\textnormal{#2}%
  \else\ifx#1pages\textnormal{#2}%
  \else\ifx#1year\textnormal{(#2)}%
  \else #2%
  \fi\fi\fi\fi
}
\makeatother

\usepackage{textcomp}
\usepackage{float}
\usepackage[percent]{overpic}
\usepackage{xspace}
\usepackage{hhline}
\usepackage{array}
\usepackage{blkarray}
\usepackage[normalem]{ulem}
\usepackage[utf8]{inputenc}
\usepackage{multirow}
\usepackage{chngcntr}
\usepackage{textgreek}
\usepackage{flushend}
\usepackage{academicons}
\usepackage{eufrak}
\usepackage{textgreek}

\definecolor{Order0}{rgb}{0.85,0.85,0.85}  
\definecolor{Order1}{rgb}{1.0,1.0,0.8}      
\definecolor{Order2}{rgb}{0.85,0.92,1.0}    
\definecolor{Order3}{rgb}{0.85,1.0,0.85}    

\DeclareFontFamily{OT1}{pzc}{}
\DeclareFontShape{OT1}{pzc}{m}{it}
{<-> s * [1.25] pzcmi7t}{}
\DeclareMathAlphabet{\mathpzc}{OT1}{pzc}
{m}{it}
\usepackage[T1]{fontenc}
\usepackage{newtxtext,newtxmath,wasysym} 

\newcommand{\ignore}[1]{}

\begin{document}


\title{Quantum-Inspired Approach to Analyzing Complex System Dynamics}

\author{P. Kafashi}
\affiliation{Department of Physics, Sharif University of Technology, Tehran 14588, Iran}

\author{M. Orujlu}
\affiliation{Department of Physics, Sharif University of Technology, Tehran 14588, Iran}

\begin{abstract}
We present a quantum information-inspired framework for analyzing complex systems through multivariate time series. In this approach the system's state is encoded into a density matrix, providing a compact  representation of higher-order correlations and dependencies. This formulation enables precise quantification of the relative influence among time series, tracking of their response to external perturbations and also the definition of a \emph{recovery timescale} without need for dimensional reduction. 
By leveraging tools such as fidelity from quantum information theory, our method naturally captures higher-order co-fluctuations beyond pairwise statistics, offering a holistic characterization of resilience and similarity in high-dimensional dynamics. We validate this approach on synthetic data generated by a 9-dimensional modified Lorenz-96 model and demonstrate its utility on real-world climate data, analyzing global temperature anomalies across nine regions, quantifying the dissimilarity of each 288-month time window up to July 2025 relative to the 1850–1874 baseline period.
\end{abstract}

\date{\today}
\maketitle

\textit{Introduction.---}Complex systems, such as biological networks, neural systems, climate dynamics, and turbulence comprise numerous interacting components whose collective behavior is nonlinear, multistable, and sensitive to perturbations \cite{Phys-Rep}. Their dynamics are influenced not only by pairwise interactions but also by high-dimensional correlations that involve multiple components simultaneously \cite{Sagdeev1988,PRX}. Typically far-from-equilibrium and governed by unobservable or high-dimensional microscale processes, these systems are best described stochastically. In this view, noise is an inherent aspect of the dynamics, and the stochastic features observed in time series reflect the system’s underlying nonlinear interactions, far beyond being artifacts of measurement. The macroscopic degrees of freedom, representing the collective behavior of many interacting components, are encoded as multivariate time series. These time series capture the system’s emergent dynamics, revealing how large-scale patterns and structures evolve over time as a result of underlying microscopic interactions.


Building on this perspective, advancements in \textit{quantum information theory} have provided powerful tools for probing high-dimensional, strongly correlated systems. These tools offer new opportunities  that extend beyond the boundaries of quantum physics. A particularly compelling direction is the application of these tools to multivariate time series, which arise in diverse areas such as neuroscience, finance, climate science, and systems biology. These datasets are often characterized by high-dimensional and nonlinear interdependencies that pose significant challenges to classical analytical techniques \cite{Sahimi2024}, thereby motivating the development of more expressive and principled frameworks \cite{whdb-2lgl}.


To address these challenges, we introduce a quantum-inspired approach that leverages tools such as density matrices and fidelity. The tools are widely used in quantum information science \cite{Nielsen2010Quantum, Wilde2013Quantum, Vedral2002Role, Horodecki2009Quantum, Giovannetti2004Quantum, Adami1997Information, cor-pic}. We demonstrate that these tools are naturally suited to analyzing complex correlations in multivariate time series. By representing the states of multivariate time series as artificial quantum states, we employ concepts such as inseparability, correlations and coherence \cite{Modi2012Classical} to reveal higher-order dependencies and dynamic behaviors in time series. We demonstrate utility of our approach on synthetic data from a modified Lorenz 96 model and apply it to global temperature anomalies across nine regions to quantify regional responses to external disturbances.\\


\textit{Methods.---}Effective analysis of a complex system necessitates the initial definition of a set of distinguishable \textit{states} that comprehensively represent the system’s dynamical properties of interest. Each state must encapsulate salient features that meaningfully differentiate the system's behavior, thereby capturing its most critical characteristics. Subsequently, a rigorous quantitative metric is required to evaluate the degree of dissimilarity or similarity between these states. Establishing both a well-defined state representation and an appropriate similarity measure enables a systematic and rigorous characterization of changes in the system's dynamics.



Consider an $N$-dimensional state variable $\mathbf{x}(t) = \bigl(x_1(t), x_2(t), \ldots, x_N(t)\bigr)$, where each component may be increasing, decreasing, or remaining approximately constant at a given time $t$. To obtain a uniform representation across heterogeneous variables, we assign to each component a binary label $a_i(t) \in \{0,1\}$, $i=1,\ldots,N$, indicating, e.g., whether $x_i(t)$ is increasing ($a_i=1$) or not increasing ($a_i=0$), relative to their previous time step. Collecting these labels into the vector
\[
\mathbf{a}(t) =\big( a_1(t), a_2(t), \ldots,a_N(t)\big)^{T} \in \{0,1\}^N,
\]
we obtain a compact encoding of the system at time $t$, with $2^N$ possible configurations. This coarse-graining discards magnitude information but ensures comparability across variables of different scales, ranges, or physical meanings. Small and large changes within the same qualitative regime are mapped to the same binary value, and the framework can be extended to multi-level encodings if more detailed qualitative states are required.

To embed this representation into a (virtual) ``Hilbert space,'' we define the virtual single-qubit space $\mathcal{H}_2 = \operatorname{span}\{\ket{0}, \ket{1}\}$, with $\ket{0} = (1,0)^{T}$ and $\ket{1}=(0,1)^{T}$. The full $N$-qubit Hilbert space is then $\mathcal{H} = \mathcal{H}_2^{\otimes N} \cong \mathds{C}^{2^N}$ , with the computational basis vectors given by $\ket{\phi(t)} = \otimes_{i=1}^N \ket{a_i(t)} = \ket{a_1(t), a_2(t), \ldots, a_N(t)}$. By construction, these $2^N$ vectors are mutually orthonormal, $\braket{\phi_k|\phi_{k'}} = \delta_{kk'}$. Interpreting the bit string $a_1 a_2 \cdots a_N$ as a binary number in big-endian order provides the 1-based index of $\ket{\phi(t)}$ in the computational basis: $\mathrm{index}(a_1 a_2 \cdots a_N) = 1 + \sum_{i=1}^N a_i \, 2^{\,N-i}$. As an example, the all-zero configuration is $\ket{00\cdots0} = \ket{0}^{\otimes N} = (1,0,\ldots,0)^{T}$.

To characterize the system's behavior over a finite observation interval $[t, t+\tau_0]$, we define the Hilbert-space state
\begin{equation}
\ket{\psi(t)} = \textstyle{\sum_{k=1}^{2^N}} \sqrt{P_k(t)} |\phi_k\rangle,
\label{coding}
\end{equation}
where $P_k(t)$ denotes the probability of observing configuration $\ket{\phi_k}$ within the interval. When multiple ensemble realizations are available, $P_k(t)$ can be estimated from their frequencies; otherwise, it can be approximated using time averages. For a single trajectory of a multivariate time series, the window size $\tau_0$ is chosen to correspond to the second-order stationarity timescale, ensuring representative sampling of the system's dynamics. 

The state $\ket{\psi(t)}$ provides a \emph{quantum-inspired embedding} of classical probabilities rather than a true quantum state. While we employ the formalism of Hilbert spaces and superposition, the underlying system is entirely classical, and no physical quantum superposition or interference occurs in that level. This framework offers a compact and structured way to encode the distribution over all possible configurations into a single vector, enabling the use of tools from quantum information theory. 
From this embedding, we construct the density matrix
\begin{equation}
\varrho(t) = |\psi(t) \rangle \langle \psi(t)|,
\end{equation}
which is a positive semidefinite matrix with trace 1, representing the probabilities of each configuration in a unified matrix form. 
By construction, the diagonal elements of the density matrix satisfy $\varrho_{kk} = P_k(t)$, directly encoding the probabilities of each configuration, whereas the offdiagonal elements $\varrho_{kk'} = \sqrt{P_k(t) P_{k'}(t)}$ depend only on the probabilities and not on the actual values of the original variables $x_i(t)$. To explore interdependencies in the Hilbert-space embedding, one can compare $\varrho(t)$ with density matrices derived from shuffled or surrogate time series, employing appropriate distance measures to detect non-random structure.

The similarity between two density matrices corresponding to consecutive observation windows provides a quantitative measure of the \emph{temporal similarity of state distributions} in the time series. Specifically, for windows $[t_1, t_1+\tau_0]$ and $[t_2, t_2+\tau_0]$, with density matrices $\varrho(t_1)$ and $\varrho(t_2)$, a suitable similarity measure such as the fidelity $F(\varrho(t_1),\varrho(t_2))$ yields a single number that captures the collective similarity between the two $N$-dimensional segments.


Unlike conventional pairwise correlation coefficients, which are limited to capturing linear relationships between individual variables, the Hilbert-space embedding retains higher than pairwise dependencies of the system without any dimensional reduction. As a result, it is intrinsically sensitive to higher-order co-fluctuations and joint dependencies among all variables, offering a substantially richer and more holistic characterization of the system’s temporal dynamics.

To analyze the temporal evolution of the system, we employ fidelity between density matrices \cite{Nielsen2010Quantum}
\begin{equation}
\label{fidelity definition}
 F(\varrho, \sigma) = \big( \textstyle{\mathrm{Tr} \sqrt{\sqrt{\varrho} \, \sigma \, \sqrt{\varrho}} }\big)^2,
\end{equation} 
which reduces to $F = |\langle \psi|\phi\rangle|^2$ for pure states $\varrho = |\psi\rangle \langle \psi|$ and $\sigma = |\phi\rangle \langle \phi|$. Fidelity ranges from $0$ to $1$, with $F=1$ for identical states and lower values indicating increasing dissimilarity. 

Classically, if the system occupies a single configuration with probability one during the observation window ($P_k = 1$ for one configuration, $P_{k'} = 0$ for all others), the embedded state reduces to a single basis vector, and the density matrix has support only on that configuration. 
More generally, when several configurations carry nonzero probability ($0 < P_k < 1$), the embedding still yields a \emph{pure state}, as in $\mathrm{Tr}[\varrho^2]=1$, yet it corresponds to a \emph{classically mixed distribution} over configurations. This distinction highlights that the purity of $\varrho(t)$ arises from the square-root embedding rather than from physical superposition.\\




\textit{Recovery Time Scale.---}In complex systems, sudden external perturbations or shifts in control parameters can induce pronounced changes in the dynamical behavior. Such responses may persist over certain timescales, relax back to the original state after a characteristic recovery period, or drive the system into qualitatively new regimes---such as alternative stable equilibria, sustained oscillations, or even chaotic trajectories.


Complex systems are inherently multidimensional, and focusing on a single observable or applying dimensional reduction can obscure collective patterns of resilience, i.e., the system’s ability to recover from perturbed states. A more rigorous approach requires analyzing multiple subsystems or observables simultaneously to capture the distributed and emergent nature of resilience in complex dynamics. To this end, we employ \emph{fidelity} between density matrices as a high-dimensional analog of classical early-warning indicators \cite{Scheffer2009}, such as increasing variance or lag-1 autocorrelation. This framework allows us to estimate recovery timescales using the full information contained in an $N$-dimensional time series. \\

\textit{Method Illustration using Synthetic and Real-World Time Series.---}To  illustrate our approach, we generate synthetic data from a representative high-dimensional dynamical system. Specifically, we employ a modified Lorenz-96 model \cite{lorenz1996predictability}, which captures complex behaviors reminiscent of real-world systems. While the standard Lorenz-96 model already exhibits rich chaotic dynamics, we introduce modifications to simulate scenarios in which the forcing on an individual component is temporarily perturbed by external stimuli and then returns to its preset value over a short time.

The Lorenz-96 model with diffusive coupling is defined as
\begin{equation}
\label{Modified Lorenz}
\frac{dx_i}{dt} = \underbrace{(x_{i+1} - x_{i-2}) x_{i-1} - x_i + F_i(t)}_{\text{Lorenz-96 dynamics}} + \underbrace{\gamma \textstyle{\sum_{j=1}^{N}} (x_j - x_i)}_{\text{diffusive coupling}}, 
\end{equation}
where $i = 1, 2, \ldots, N$, $\gamma$ is the coupling constant, and $F_i(t)$ denotes the external forcing applied to the component $x_i$. A common choice is a constant forcing $F_i(t) = 8$, for which the system exhibits high-dimensional chaotic behavior when uncoupled ($\gamma = 0$) \cite{10.1063/1.3496397}.  

In Eq. (\ref{Modified Lorenz}) each $x_i$ interacts with other variables through two mechanisms: (i) the nonlinear term inherited from the original Lorenz-96 dynamics, and (ii) an additional diffusive coupling term proportional to $(x_j - x_i)$, which promotes synchronization among components.  
As $\gamma$ increases, the influence of surrounding states grows, driving each $x_i$ closer to the average behavior of the coupled variables and enhancing collective coherence across the system.

\begin{figure}[tp]
\includegraphics[width=\linewidth]{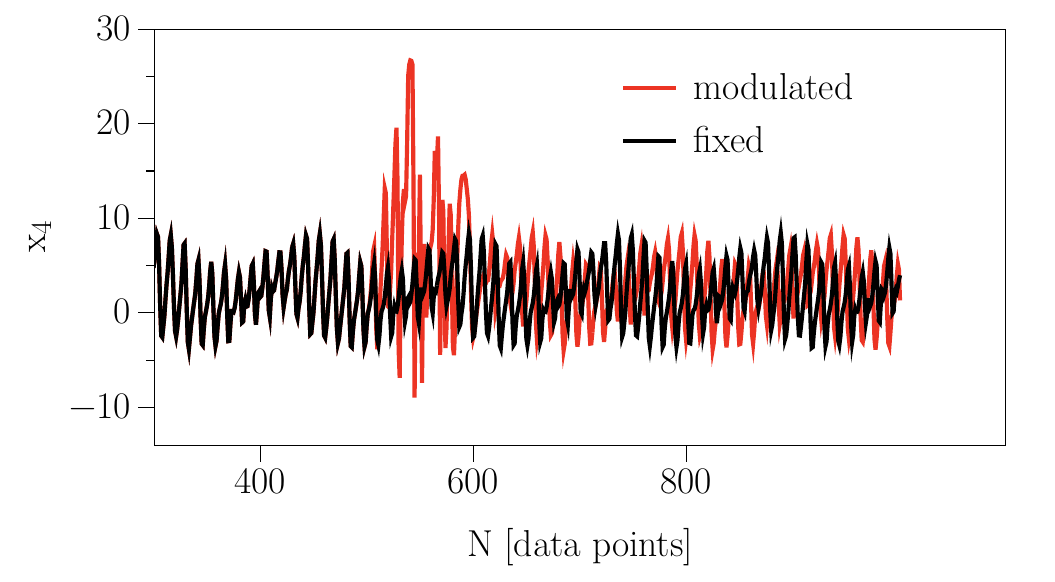}
\vspace{0.4cm}
\includegraphics[width=\linewidth]{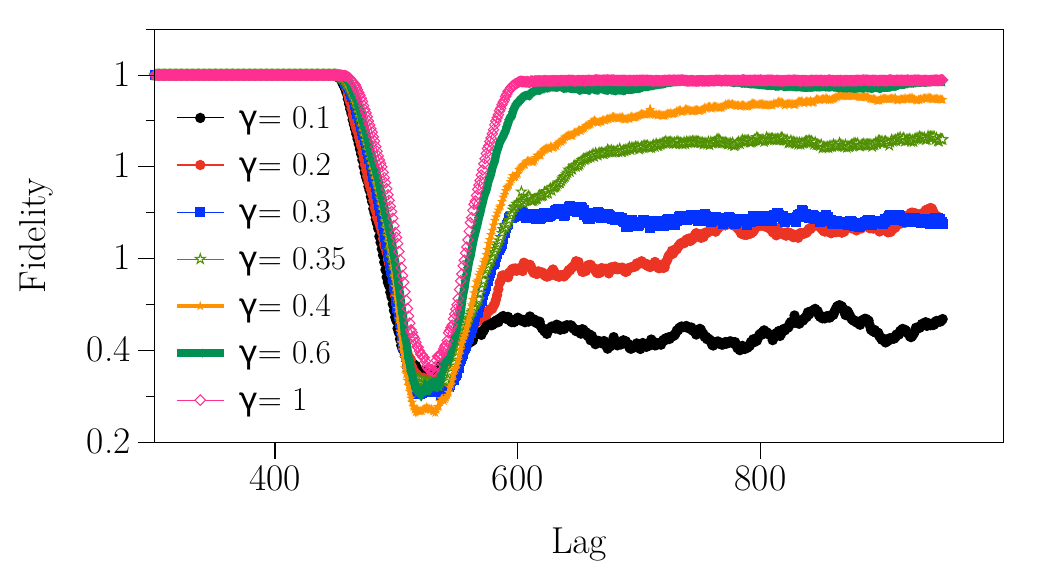}
\vspace{-0.9cm}
\caption{(a) Perturbation applied to the fourth component of the model: the forcing parameter $F_4$ increases from 8 to 90 over 50 time steps starting at $N=500$, and then decreases back to 8 over the following 50 steps. (b) Fidelity between a system with constant forcing (fixed) and one with a sudden spike in the forcing parameter (modulated), shown for different values of $\gamma$ and averaged over 30 independent realizations with different initial conditions. Larger coupling strengths $\gamma$ lead to faster recovery.}
\label{fig1}
\end{figure}

We compare two 9-dimensional Lorenz-96 systems simulated with a time step of $\dt = 0.1$. The first system evolves under constant forcing, while in the second, a sudden spike is applied to the forcing of component $x_4$ ($F_4$) at $N = 500$ (Fig.~\ref{fig1}a). Both systems are initialized with the same random seed to ensure a consistent basis for comparison. To quantify their divergence, we construct windowed density matrices, $\varrho(t)$ and $\sigma(t)$, using segments of 50 data points. Within each window, we estimate the probability distributions $P_k$ for each system, which are then used to assemble the corresponding density matrices. The fidelity between $\varrho(t)$ and $\sigma(t)$, computed via Eq.~\eqref{fidelity definition}, provides a quantitative measure of the similarity between the perturbed and unperturbed dynamics.
 

  In Fig.~\ref{fig1}b, the fidelities clearly illustrate the impact of the coupling strength. For larger values of $\gamma$, the perturbed system almost fully returns to its original state, whereas for smaller $\gamma$, recovery remains limited. This highlights the critical role of the diffusive coupling in stabilizing dynamics and preserving the system’s intrinsic state. At the same time, it demonstrates the sensitivity of our fidelity-based framework in detecting and quantifying recovery time scale. We remark that the use of fidelity as a recovery measure has also been emphasized in other related works \cite{7404264, PhysRevA.92.042321}.\\

\textit{Implementing on a Real Data Set.---}We apply our approach to real monthly climate records from the NOAA Global Time Series dataset, spanning January 1850 to July 2025 \cite{noaa_climate_gla_2025}. The dataset contains temperature anomalies, deviations from a long-term average rather than absolute temperatures. For all regions considered here, anomalies are computed relative to the 1910--2000 mean and expressed in degrees Celsius. Our analysis encompasses nine distinct geographical regions: North America, Europe, the Arctic, Asia, Antarctica, Oceania, South America, Africa, and the Atlantic Meridional Overturning Circulation (MDR) region.  The average second-order stationarity of the time series was assessed using the Augmented Dickey-Fuller test, which indicated a window size ranging from approximately 60 to 288 data points with $p < 0.05$. Based on this, we used a window size of 288 for all subsequent analyses.

For this analysis, a reference state needs to be defined as a neutral baseline against which all other states are compared. To identify regions affected by changes in temperature patterns, we take the period preceding the second Industrial Revolution as the reference, representing a state largely free from anthropogenic influences. Accordingly, we use the first window of 288 months, beginning in January 1850, as our reference state, which is second-order stationary.

\begin{figure}[tp]
    \includegraphics[width=\linewidth]{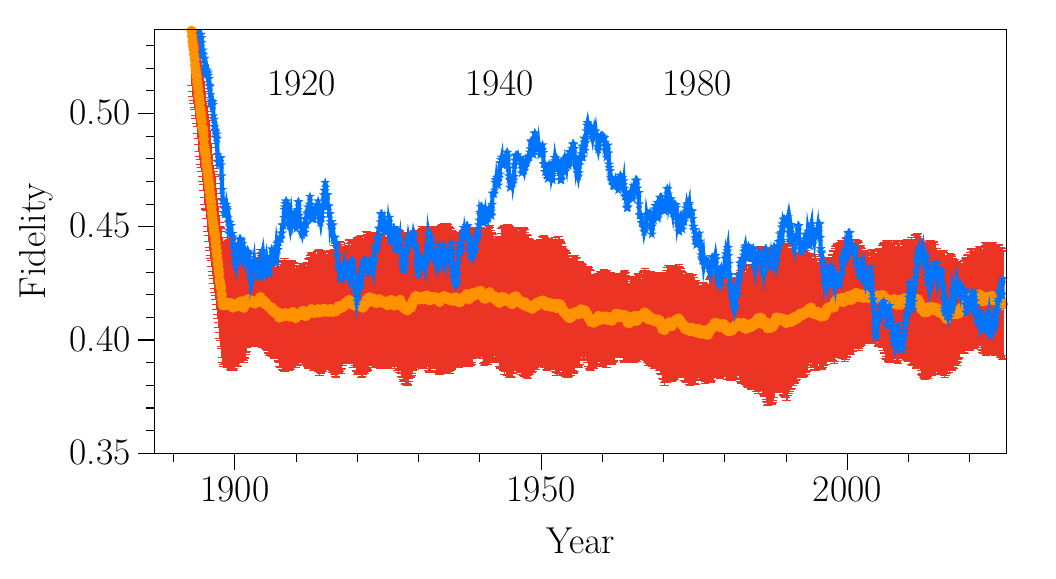}
%
%
\vspace{-0.7cm}
\caption{The fidelity of the reference density matrix from the first window compared with the density matrix at each subsequent time, as a function of time. The blue line shows the fidelity for the system of 9 regional temperature series from January 1850 to July 2025, while the orange line corresponds to the fidelity for shuffled (white-noise) time series of the same window length. The shaded red region denotes the standard deviation across 20 shuffled realizations.  
}
 \label{Full fidelity Temperature}
\end{figure}

The total system fidelity, along with the average fidelity computed from systems driven by shuffled (white-noise) data, is shown in Fig. \ref{Full fidelity Temperature}. To calculate this, we used a moving window of 288 months starting in January 1850. At each step, the window was advanced by one month, the corresponding density matrix was recalculated, and the fidelity with respect to the reference state (the first 288 months) was evaluated. This procedure tracks the system’s evolution over time relative to the baseline and allows comparison with the expected fidelity of purely random (shuffled) data, providing a benchmark for identifying system-specific dynamics. Two distinct intervals, 1910-1920 and 1940-1980, exhibit the highest similarity to the initial 288 months, while after 1980 no significant similarity is observed.
 



The tools introduced above allow subsystem-level analysis while preserving interrelations in the full system, enabling us to isolate the contribution of each region. To identify which regions differ most from the reference window (288 months starting in January 1850), we use the reduction via partial tracing, focusing on subsystems $A$ of size 1 (individual regions). The reduced density matrix is obtained by tracing out all other components, $\varrho_A(t) = \mathrm{Tr}_B[\varrho(t)]$, where $B$ is the complement of $A$.   The \emph{subsystem fidelity} is defined as
\begin{equation}
f_A(t) = F\big(\varrho_A(0), \varrho_A(t)\big),
\end{equation}
measuring the similarity between a subsystem’s initial state and its state at time $t$. For subsystems of size 1, each $f_i(t)$ quantifies how closely region $i$ follows its initial dynamics. 
 
\begin{figure}[tp]
\includegraphics[width=0.9\linewidth]{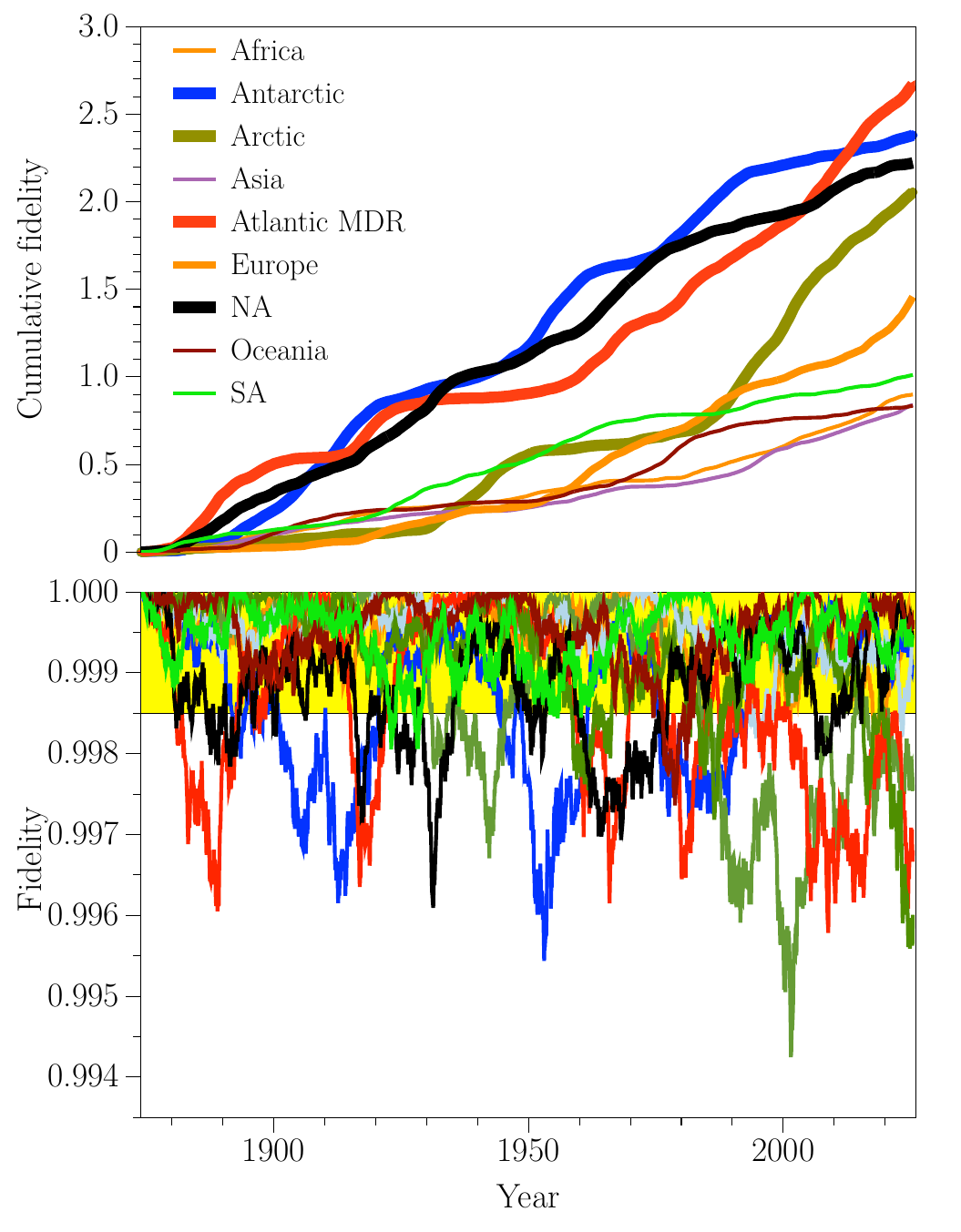}
\vskip-3mm
\caption{Bottom: Subsystem fidelity between the reference density matrix (first window) and subsequent windows as a function of time. Each colored line corresponds to a subsystem of size 1 for a given region. The yellow shaded band shows the standard deviation of the fidelity for a white-noise system of equal size and window length, averaged over 20 realizations. Top: Cumulative deviation of fidelity from $y=1$ for the subsystems shown below. At each time $t$, the integrated area between the fidelity curve and $y=1$, from the beginning up to $t$, is plotted as the cumulative area.}
\label{fig:placeholder}
\end{figure}

In the bottom panel of Fig. \ref{fig:placeholder}, the fidelities for subsystems of size 1 are shown, along with the average fidelity for systems driven by white noise (yellow shaded area). Fidelity values that fall outside this region indicate that the system is exhibiting behavior distinct from random noise.


From the bottom panel of Fig. \ref{fig:placeholder}, we can examine each region individually to assess its deviation relative to the pre-industrial era. To quantify this, we define the \emph{cumulative fidelity} as the area between the fidelity curve (bottom panel of Fig. \ref{fig:placeholder}) and the line $y = 1$, integrated from the initial time up to time $t$. This measure is displayed as a function of time in the top panel of Fig. \ref{fig:placeholder}. The cumulative fidelity provides a clear metric for evaluating the overall impact on each subsystem. While alternative definitions could be employed depending on the specific goal, here we use it to identify the regions that have experienced the greatest total change over time.

As shown in Fig. \ref{fig:placeholder}, four regions- the Atlantic MDR, Antarctic, North America, and Arctic- display the largest integrated deviations from the 1850 baseline. Notably, the Arctic exhibits a pronounced increase beginning around 1970. In recent years, its rate of change closely parallels that of the Atlantic MDR, indicating that both the Arctic and Atlantic MDR are becoming the regions most strongly diverging from the 1850 baseline conditions.
Similarly, Europe has begun to show a significant increase in dissimilarity. 

Importantly, our framework is fully compatible with the mathematical formalism of quantum mechanics. Each system state is represented as a density matrix---a positive semidefinite, trace-one Hermitian operator--which encodes the probabilities of all configurations within a single object. This construction ensures that standard quantum information measures, such as fidelity and entropy, are well defined and directly applicable. Moreover, subsystem dynamics can be analyzed via partial traces, and recovery or similarity measures can be interpreted in terms of quantum-state distinguishability. While the underlying system is classical, the formal structure mirrors that of genuine quantum states, making the framework naturally mappable to quantum computing architectures for potential implementation on qubit-based processors.

By bridging techniques from quantum information theory and time series analysis, a compact and principled way is provided to probe the structure and evolution of complex systems. This cross-disciplinary, quantum-inspired approach, drawing on ideas from information theory and data science, opens new avenues for both theoretical investigations and practical applications~\cite{mugel2020dynamic, PhysRevA.111.032409, emmanoulopoulos2022quantum}.\\

\textit{Acknowledgments}.---P.K. and M.O. gratefully acknowledge M. R. Rahimi Tabar and M. Asoudeh for insightful discussions and valuable feedback. \\

\textit{Code availability.---}All codes developed in this study are publicly available \cite{code}.


\bibliography{refs.bib}

@misc{code,
  author       = {Kafashi, P.},
  title        = {Quantum-Compatible-Approach-to-Complex-System-Dynamics-Codes},
  year         = {2025},
  url          = {https://github.com/PaRaShi12/Quantum-Compatible-Approach-to-Complex-System-Dynamics-Codes-},
  note         = {GitHub repository}
}

@article{PRX,
  title = {Revealing Higher-Order Interactions in High-Dimensional Complex Systems: {A} Data-Driven Approach},
  author = {Tabar, M. Reza Rahimi and Nikakhtar, Farnik and Parkavousi, Laya and Akhshi, Amin and Feudel, Ulrike and Lehnertz, Klaus},
  journal = {Phys. Rev. X},
  volume = {14},
  issue = {1},
  pages = {011050},
  numpages = {36},
  year = {2024},
  doi = {10.1103/PhysRevX.14.011050}
}

@Article{Phys-Rep,
  author  = {Friedrich, Rudolf and Peinke, Joachim and Sahimi, Muhammad and Tabar, M Reza Rahimi},
  journal = {Phys. Rep.},
  title   = {Approaching complexity by stochastic methods: {From} biological systems to turbulence},
  year    = {2011},
  number  = {5},
  pages   = {87},
  volume  = {506},
  doi     = {10.1016/j.physrep.2011.05.003}
}

@article{Sahimi2024,
  title = {Physics-informed and data-driven discovery of governing equations for complex phenomena in heterogeneous media},
  author = {Sahimi, Muhammad},
  journal = {Phys. Rev. E},
  volume = {109},
  issue = {4},
  pages = {041001},
  year = {2024},
  doi = {10.1103/PhysRevE.109.041001}
}

@inproceedings{lorenz1996predictability,
  author    = {Edward N. Lorenz},
  title     = {Predictability: A Problem Partly Solved},
  booktitle = {Predictability of Weather and Climate},
editor = {Palmer, T. and Hagedorn, R. },
  year      = {2009},
Publisher = {Cambridge University Press},
Address = {Cambridge, UK},
  pages     = {1--18}
}

@book{Nielsen2010Quantum,
  title     = {Quantum Computation and Quantum Information},
  author    = {Nielsen, Michael A. and Chuang, Isaac L.},
  year      = {2010},
  publisher = {Cambridge University Press},
Address = {Cambridge, UK}
}

@book{Wilde2013Quantum,
  title     = {Quantum Information Theory},
  author    = {Wilde, Mark M.},
  year      = {2013},
  publisher = {Cambridge University Press},
Address = {Cambridge, UK}
}

@article{Modi2012Classical,
  author  = {Modi, Kavan and Brodutch, Aharon and Cable, Hugo and Paterek, Tomasz and Vedral, Vlatko},
  title   = {The classical-quantum boundary for correlations: {Discord} and related measures},
  journal = {Rev. Mod. Phys.},
  volume  = {84},
  number  = {4},
  pages   = {1655},
  year    = {2012},
Doi = {10.1103/RevModPhys.84.1655}
}

@article{Vedral2002Role,
  author  = {Vedral, Vlatko},
  title   = {The role of relative entropy in quantum information theory},
  journal = {Rev. Mod. Phys.},
  volume  = {74},
  number  = {1},
  pages   = {197},
  year    = {2002},
Doi = {10.1103/RevModPhys.74.197}
}

@article{Horodecki2009Quantum,
  author  = {Horodecki, R. and Horodecki, P. and Horodecki, M. and Horodecki, K.},
  title   = {Quantum entanglement},
  journal = {Rev. Mod. Phys.},
  volume  = {81},
  number  = {2},
  pages   = {865},
  year    = {2009},
Doi = {10.1103/RevModPhys.81.865}
}

@article{Giovannetti2004Quantum,
  author  = {Giovannetti, V. and Lloyd, S. and Maccone, L.},
  title   = {Quantum-enhanced measurements: beating the standard quantum limit},
  journal = {Science},
  volume  = {306},
  number  = {5700},
  pages   = {1330},
  year    = {2004},
Doi = {10.1126/science.1104149}
}

@article{Adami1997Information,
  author  = {Adami, C. and Cerf, N. J.},
  title   = {{von Neumann} capacity of noisy quantum channels},
  journal = {Phys. Rev. A},
  volume  = {56},
  number  = {5},
  pages   = {3470},
  year    = {1997},
Doi = {10.1103/PhysRevA.56.3470}
}

@book{Sagdeev1988,
  author       = {R. Z. Sagdeev and D. A. Usikov and G. M. Zaslavsky},
  title        = {Nonlinear Physics: From the Pendulum to Turbulence and Chaos},
  year         = {1988},
  publisher    = {Harwood Academic Publishers},
  address      = {Chur, Switzerland}
}

@misc{emmanoulopoulos2022quantum,
      title={Quantum Machine Learning in Finance: {Time} Series Forecasting}, 
      author={Dimitrios Emmanoulopoulos and Sofija Dimoska},
      eprint={2202.00599},
      archivePrefix={arXiv},
      doi={10.48550/arXiv.2202.00599} 
}

@article{mugel2020dynamic,
  title = {Dynamic Portfolio Optimization with Real Datasets Using Quantum Processors and Quantum-Inspired Tensor Networks},
  author = {Mugel, Samuel and Kuchkovsky, Carlos and Sánchez, Escolástico and Fern{\'a}ndez-Lorenzo, Samuel and Luis-Hita, Jorge and Lizaso, Enrique and Or{\'u}s, Román},
  journal = {Phys. Rev. Research},
  volume = {4},
  Pages = {013006},
  year = {2022},
Doi = {10.1103/PhysRevResearch.4.013006}
}

@article{PhysRevA.111.032409,
  title = {Tensor networks and efficient descriptions of classical data},
  author = {Lu, Sirui and Kan\'asz-Nagy, M\'arton and Kukuljan, Ivan and Cirac, J. Ignacio},
  journal = {Phys. Rev. A},
  volume = {111},
  issue = {3},
  pages = {032409},
  year = {2025},
  doi = {10.1103/PhysRevA.111.032409}
}

@article{whdb-2lgl,
  title = {Effective One-Dimensional Reduction of Multicompartment Complex Systems Dynamics},
  author = {Visco, Giorgio Vittorio and Nauta, Johannes and Scagliarini, Tomas and Artime, Oriol and De Domenico, Manlio},
  journal = {Phys. Rev. X},
  volume = {15},
  issue = {3},
  pages = {031017},
  year = {2025},
  doi = {10.1103/whdb-2lgl}
}

@ARTICLE{7404264,
  author={Berta, Mario and Tomamichel, Marco},
  journal={IEEE Trans. Inf. Theo.}, 
  title={The Fidelity of Recovery Is Multiplicative}, 
  year={2016},
  volume={62},
  number={4},
  pages={1758},
  doi={10.1109/TIT.2016.2527683}
}

@article{PhysRevA.92.042321,
  title = {Fidelity of recovery, squashed entanglement, and measurement recoverability},
  author = {Seshadreesan, Kaushik P. and Wilde, Mark M.},
  journal = {Phys. Rev. A},
  volume = {92},
  issue = {4},
  pages = {042321},
  numpages = {23},
  year = {2015},
  doi = {10.1103/PhysRevA.92.042321}
}

@misc{noaa_climate_gla_2025,
Author = "{NOAA National Centers for Environmental Information}",
  title        = {Climate at a glance: {Global} time series},
  year         = {2025},
  Url = {https://www.ncei.noaa.gov/access/monitoring/climate-at-a-glance/global/time-series}
}

@article{10.1063/1.3496397,
    author = {Karimi, A. and Paul, M. R.},
    title = {Extensive chaos in the Lorenz-96 model},
    journal = {Chaos},
    volume = {20},
    number = {4},
    pages = {043105},
    year = {2010},
 doi = {10.1063/1.3496397}
}

@article{Scheffer2009,
  title   = {Early-warning signals for critical transitions},
  author  = {Scheffer, Marten and Bascompte, Jordi and Brock, William A. and Brovkin, Victor and Carpenter, Stephen R. and Dakos, Vasilis and Held, Hermann and van Nes, Egbert H. and Rietkerk, Max and Sugihara, George},
  journal = {Nature},
  volume  = {461},
  number  = {7260},
  pages   = {53--59},
  year    = {2009},
  doi     = {10.1038/nature08227},
  url     = {https://doi.org/10.1038/nature08227}
}

@article{cor-pic,
  title = {Correlation-Picture Approach to Open-Quantum-System Dynamics},
  author = {Alipour, S. and Rezakhani, A. T. and Babu, A. P. and M\o{}lmer, K. and M\"ott\"onen, M. and Ala-Nissila, T.},
  journal = {Phys. Rev. X},
  volume = {10},
  issue = {4},
  pages = {041024},
  year = {2020},
  doi = {10.1103/PhysRevX.10.041024}
}

\end{document}